\documentclass[twocolumn,showpacs,preprintnumbers,amsmath,amssymb,pre]{revtex4}
\usepackage{times}
\usepackage[sort&compress]{natbib}
\usepackage[dvips]{graphicx}

\usepackage{color}

\begin{document}

\preprint{}

\title{Finite population size effects in quasispecies models with single-peak fitness landscape }

\author{David B. Saakian$^{1,2,3}$}
\email{saakian@yerphi.am}
\author{Michael W. Deem$^4$ }
\author{Chin-Kun Hu$^2$}
\email{huck@phys.sinica.edu.tw}
 \affiliation{$^1$Yerevan Physics Institute, 2 Alikhanian
Brothers St, Yerevan 375036, Armenia}
 \affiliation{$^2$Institute of Physics, Academia Sinica, Nankang, Taipei 11529, Taiwan}
 \affiliation{$^3$ National Center for Theoretical Sciences: Physics
Division, National Taiwan University, Taipei 106, Taiwan}
 \affiliation{$^4$ Department of Physics \& Astronomy, Rice
University, Houston,Texas 77005--1892, USA }

\begin{abstract}
We consider finite population size effects for Crow-Kimura and Eigen
quasispecies models with single peak fitness landscape. We formulate
accurately the iteration procedure for the finite population models,
then derive Hamilton-Jacobi equation (HJE) to describe the dynamic
of the probability distribution. The steady state solution of HJE
gives the variance of the mean fitness. Our results are useful for
understanding population sizes of virus in which the infinite
population models can give reliable results for the biological
evolution problems. \end{abstract}

 \pacs{87.23.Kg,87.10.Mn}

\maketitle


\maketitle


\section{Introduction}

Investigation of biological evolution models
\cite{wr32,mo,ei71,ei89,ck70,ew04}, such as Eigen model
\cite{ei71,ei89} and Crow-Kimura model \cite{ck70}, by methods of
statistical or theoretical physics is highly fruitful in evolution
research. The methods used include quantum mechanics
\cite{ba97,bw01}, statistical mechanics \cite{sa04}-\cite{sa06},
quantum field theory \cite{sa04}-\cite{de07},\cite{de08},
Hamilton-Jacobi equation (HJE) \cite{sa07,ka07,sa08a}. Such approach
has given many exact results for evolution models
\cite{ei71}-\cite{hi96}, solved a paradox of the origin of life
\cite{11PlosOne}, and produced exact finite genome length
corrections for the mean fitness and gene probabilities in some
evolution models \cite{11jsp-fsc}.

In the original formulation of the Eigen and Crow-Kimura models, the
configurations of the genome of length $L$ are represented by $M
\equiv 2^L$ spin configurations $(s_1,s_2,\dots,s_L)$, where $s_k$
for $1 \le k \le L$ take $+1$ or $-1$. Such representation was used
by Peng, {\it et al.} to study long-range correlation in nucleotide
sequences \cite{92natureDNA}. The $M$ configurations $S_i$ are
labelled by $0 \le i \le M-1$ and the $i-$th configuration $S_i$ is
assigned a number $r_i$ to represent the reproduction rate or
fitness of that configuration and another number $p_i$ to represent
the probability in that configurations. Such $p_i$ satisfy the
normalization condition: $\sum_{i=0}^{M-1} p_i=1$. The coupled
differential equations satisfied by $p_i$ for the Crow-Kimura model
\cite{ck70} and Eigen model \cite{ei71,ei89} are given in Appendix A
and Appendix B, respectively. However, such coupled differential
equations are valid only in the limit of infinite population size
$N$, which is not the case in many real systems, e.g. virus in a
given environment. Thus the study of finite population size problem
has attracted much attention in recent decades
\cite{ns89}-\cite{kr10}. While the case of two alleles (types of
genes) in the Wright-Fisher model \cite{wr32} and the Moran model
\cite{mo} can be analytically solved \cite{ew04}, the realistic case
of evolution with many sequences (genomes) stays intractable by
traditional methods. In \cite{kr00} the additive fitness landscape
has been considered, when the contributions of different alleles to
the fitness are random numbers and in \cite{kr07} a finite
population was considered in which  the finesses of different
sequences are independent random numbers.

The purpose of the present Letter is to formulate Crow-Kimura model
and Eigen model for finite population size and solve them for the
single peak fitness landscape, popular in quasispecies literature.
In such a landscape, the fitness of a configuration, say $S_0$, is
larger than fitness of other configurations, i.e. $ r_0 >  r_i$ for
$i >0$, and all $r_i$ are equal for $i\ge 1$. We first formulate the
iteration procedure for the finite population models, then derive
HJE to describe the dynamic of the probability distribution. The
steady state solution of HJE gives the variance of the mean fitness.
Our results are useful for understanding population sizes of virus
in which the coupled differential equations can give reliable
results for biological evolution problems. Our results are exact
derivations, versus numerics or uncontrolled approximations in vast
majority of finite population articles.



Consider the case when the total number of different genotypes $N_g$
is either $N_g\sim 4^L$, where $L$ is the number of nucleotides in
the virus genome, or $N_g\sim 2^L$, where $L$ is the number of two
type of alleles  or (spins), located in different places (loci). The
infinite population case is when the population size $N$ is large
enough to have large number of viruses of any type. The convergence
of evolutionary dynamics with population size
 depends on the mutation rate and the fitness landscape.
In the infinite population limit the evolution equations are
deterministic, and, for molecular evolution models
\cite{ei71}-\cite{ba97}, there are many exact results
\cite{ba97}-\cite{de08}. It is possible even to find exact solutions
for the steady state and dynamics \cite{sa07,ka07,sa08}. In biology,
the populations are often relatively small.  Then the collective
characteristics of the evolving population will fluctuate.


\section{Finite population Crow-Kimura model with symmetric fitness
landscape}

We consider the symmetric fitness landscape, popular in virology.
The genome is a collection of letters (spins) $\pm 1$, denoting the
alleles type. Thus a sequence is identified with a spin
configuration of a one-dimensional Ising model. By mutation any
letter may randomly change to the other value.

An important characteristics of the  sequence space is the number of
neighbors $L$ and the number of sequences at the Hamming distance
$d$ (sequences have $d$ alleles different from the reference
sequence) $N_d=\frac{L!}{d!(L-d)!}\sim \frac{L^d}{d!}$ for two loci
case at $d\ll L$. We consider a simple fitness landscape, popular in
population genetics, when the sequence from the $l$-th Hamming class
has a fitness $w_l$.

As evolution is a stochastic process, we should work with
probabilities. We are interested in steady state properties of the
evolution model.  We consider the finite population version of
Crow-Kimura model, see the Appendix.  In this case we consider the
evolution as a Markov process, where the state (the state of the
population is characterized by the number of individuals for the
different possible numbers of mutations)
 is defined by $L+1$
integers, the numbers of different Hamming types.
 During one
evolution step, there are three processes: birth of new individuals
proportional to the fitness of the corresponding Hamming class,
transitions between Hamming classes proportional to $l/L$ to the
lower Hamming class and transitions proportional to $(L-l)/L$ to the
higher Hamming class \cite{bw01,hi96}. These factors $l/L$ and
$(L-l)/L$ have been derived in \cite{bw01,hi96} for the infinite
population models, and should be applied to the discrete time scheme
of the finite population models as well. The iteration step is
completed by the random reduction of the population to the initial
size, $N$. This evolution dynamics described here corresponds to the
Moran model with many alleles. Compared to the ordinary multi-allele
Moran model \cite{mo}, in our case there is a non-trivial geometry
in sequence space, defined by Hamming distance.

We first define our model for the case of a general symmetric
fitness landscape with Wrightian fitness  $\hat r_l=e^{U r_l},$ in
the $l$-th Hamming class, $0\le l\le L$, where $r_l$ is the fitness
defined earlier. Here the average number of mutations of genome per
one replication period is $U\equiv \gamma_0\tau$, where $\gamma_0$
is a mutation rate per genome in the continuous-time parallel model,
and $\tau$ is the time step. At any moment the state of our model is
characterized via $L+1$ integers $n_l$. We choose $\gamma_0=1$,
therefore $U=\tau$. We consider the $U\ll 1$ limit. In this case the
steady state results and dynamics are $U$ independent (U gives just
the scale).

During the iteration step we consider the following processes:
\begin {itemize}

\item A. Random growth with $n_l\to n_l+\delta n_l$. The $\delta
n_l$ is a random binomial process with probability $U \hat r_l$ and
$n_l$ trials.

  \item B. Mutations.

  There are $f_l$ forward mutations from the class {\it l}.

  We consider random integer number $f_l$ with binomial distribution with the probability parameter $(1-\frac{l}{L})U$
  and  $n_l$ trials.
  There are $b_l$ back mutations from the class {\it l}.

    We consider random integer number $b_l$ with binomial distribution with  the probability parameter $\frac{l}{L}U$
  and  $n_l$ trials.
Due to back mutations we have the following change of $n_l$: $n_l\to
n_l-b_l+b_{l+1}$.

  \item C.
   We randomly remove $\sum_l \delta n_l$ individuals from the
   population to keep a fixed population size.
  \end{itemize}

 \section{Sharp peak (single peak) model}

Consider the Wrightian fitness  with $r_0=e^{\epsilon J}$ for the
peak sequence,  $U<<1$ is the number of mutations per generation,
$J$ is a fitness gap in the corresponding continuous-time parallel
selection-mutation model, and $r_i=1$ for $i\ge 1$. Our goal is to
investigate how the mean fitness depends on finite population size.

 In case of infinite population, one can calculate
the number of viruses with the peak sequence using a single
equation, with the $1/L$ accuracy. Assume that there is some
probability distribution $\rho(n)$ for the number $n$ of viruses
with the peak sequence, which satisfies the normalization condition
$\sum _{n=0}^N \rho(n)=1$. Then we can derive both the steady state
distribution, which is a rather simple function, and even the exact
dynamics, which is a complicated expression for the $\rho(n)$.

We consider a discrete time scheme of evolution with small $U$.
During each iteration we consider the steps A, B, C. In Step A,
there are $\delta n$ new viruses at the peak sequence. The number
$\delta n$ is derived  via a binomial $n$ sampling with a small
probability $U J$. During the step C of reduction to keep a constant
population size anyone of this  $\delta n$ viruses could be removed
from the system. The total number of removed viruses from the peak
sequence $m$ is calculated via binomial distribution with  $\delta n
$ trials and the probability $x\equiv n/N$. Therefore, the result of
A and C steps should be a sampling of $n$ particles with a
probability $U J(1-x)$. Thus after steps A and C the original $n$
changes as $n\to n+h$, where $h$ has a binomial distribution with a
probability parameter $p=U J(1-x)$, and the number of trials is $n$.
During the step B of mutation, $n\to n-m$, where $m$ has a binomial
distribution with a probability parameter $U$ and the number of
trials is $n$. Thus after one iteration $n\to n+h-m$.

If we have a distribution $\rho(t,n)$ at the $t$-th moment of time,
then after an iteration with the period of time $U$ we have a
distribution:
\begin{equation}
\label{e1}
 \rho(t+U,n)=<\rho(t,n-h+m)>
\end{equation}
when the averaging is over the (binomial) distributions of $h$ and
$m$.

Let us assume the following anzats for the probability distribution
at the time $t$:
\begin{equation}
\label{e2}
 \rho(t,n)=\exp[N \phi(t,x)],~ x=n/N.
\end{equation}
 After an iteration
\begin{equation}
\label{e3}
 e^{N \phi(t+U,x)}=\int d t e^{N\phi(x)}<e^{-(h-m)\phi'
 (t,x)}>|_{h,m},
\end{equation}
where $\phi'\equiv \frac{\partial \phi(t,x)}{\partial x}$.
 As we used binomial probability
distributions in the iteration step, we should perform an average
via the binomial distribution in Eq. (3).
 We use the following
formula of the binomial distribution of the $h$ with a success
probability $p$ and $M$ trials:
\begin{eqnarray}
\label{e4} <e^{hk}>&& \equiv
\sum_{h=0}^{M}e^{hk}p^h(1-p)^{M-h}\frac{M!}{h!(M-h)!}\nonumber\\
&&=(1+p(e^k-1))^{M},\nonumber\\
&&\approx e^{pM(e^k-1)}.
\end{eqnarray}
We consider the case of small $p\ll 1$.

Taking $k=-\phi',M=Nx$ and $p=U J(1-x)$ in Eq.(4) (see the
definition of iteration steps A,C) we find
\begin{eqnarray}
\label{e5} <e^{-h\phi'}>=[(1-x)U Je^{-\phi'}+1-U J(1-
x)]^{Nx}\nonumber\\
\approx e^{NUJx(1-x)(e^{-\phi'}-1)}.
\end{eqnarray}

In the same way we consider the mutation, taking $k=\phi',p=U,M=xN$
we derive
\begin{eqnarray}
\label{e6} <e^{\phi'm}>=[U e^{\phi'}+1-U]^{x N}=\exp[NU
x(e^{\phi'}-1)].
\end{eqnarray}
Combining Eqs.(5),(6) and holding only the linear terms in $U$, we
obtain the following expression
\begin{equation}
\label{e7}
 \phi(t+U,x)= \phi(t,x)+UxJ(1-x)(e^{-\phi'}-1)+U x(e^{\phi'}-1)
\end{equation}
or
\begin{equation}
\label{e8}
 \frac{\partial \phi(t,x)}{\partial t}=xJ(1-x)(e^{-\phi'}-1)+x(e^{\phi'}-1).
\end{equation}

 In the steady state we just have an ordinary differential equation for $\phi$.
 We derive the following nontrivial solution $\phi_0(x)=\phi(\infty,x)$
 and the corresponding distribution
\begin{eqnarray}
\label{e9} \phi_0(x)=\int_{x_0}^xdx  \ln J(1-x)=
(x-x_0)\ln J+\nonumber\\
  (1-x)(1-\ln (1-x))-(1-x_0)(1-\ln (1-x_0)), \nonumber\\
 \rho(x)=\sqrt{\frac{NJ}{2\pi}}\exp[N\phi(x)],
\end{eqnarray}
where we added the pre-factor $\sqrt{\frac{NJ}{2\pi}}$ to ensure the
condition that total probability is 1.  Here the distribution has a
maximum at $x=x_0\equiv (1-1/J)$, see \cite{sa04}, and
 $\phi(x_0)''=-J$.
 Then
 we derive
 for the variance:
\begin{eqnarray}
\label{e10}
 V\equiv (<x^2>-<x>^2)N=\frac{1}{J}.
\end{eqnarray}
 Thus we derived
the whole steady state distribution via Eqs. (\ref{e2}),(\ref{e8}),
and the expression for the variance Eq.(\ref{e10}).

Equation (\ref{e10}) is verified numerically in Fig. 1 for
$J=1.5,~2,~3,~4$.  One could follow the method used in \cite{sa08}
to solve Eq. (\ref{e8}) and get time evolution of $\phi(t,x)$.



\section{Finite population version of the Eigen model}

Consider now the finite population version of the Eigen  model with
zero degradation. There are $n$ viruses at the peak sequence.

At any discrete moment of time we consider three processes:\\
A. the number of viruses in the class $l$ grows with a probability
$U r_l$ There are mutations. New viruses mutate with a finite
mutation probability $1-Q$,\\
C. There is a dilution of the whole population, keeping strictly the
total population size as $N$.

Consider again the single peak fitness, $ r_0=A$, and for $l>0,~
r_l=1$.  $n$ is the number of the viruses with the peak sequence,
and $x=n/N$.

\vskip 2 mm

Let us give the details of the processes A and B.

\vskip 2 mm

\noindent
 A1. Reproduction in the peak sequence $S_0$:
 We randomly choose $l$ elements from a pool of $n$ elements and each element
 is chosen independently with a probability
$U A$. Thus the probability to get $l$ elements is
\begin{eqnarray}
\label{e11}\rho_1(l)=\frac{n!}{l!(n-l)!}(U A)^l(1-U A)^{n-l}.
\end{eqnarray}
 $l$ is the number of new sequences at the peak sequence.

\noindent A2. Reproduction in the other sequences, i.e. $S_i$ for $i
>0$: We randomly choose $k$ elements from a pool of $(N-n)$ elements
and each element is chosen independently with the probability $U$.
Thus the probability distribution to get $k$ elements is
\begin{eqnarray}
\label{e12}\rho_2(k)=\frac{(N-n)!}{k!(N-n-k)!}U^k(1-U)^{N-n-k}.
\end{eqnarray}
     After A1 and A2 steps there are $n+l$ viruses at the peak
     sequences and $N-n+k$ sequences at other sequences.

     \vskip 2 mm

\noindent
 B. We randomly choose $m$ elements from a pool of $l$ elements in $S_0$ and each element is chosen
 independently with the probability $Q=\exp[-\gamma]$ to be in $S_0$. Thus the
probability to get $m$ elements in $S_0$ is:
\begin{eqnarray}
\label{e13}\rho_3(m)=\frac{l!}{m!(l-m)!}Q^m(1-Q)^{l-m}.
\end{eqnarray}
 After the step B, there are $n+m$ viruses in the peak
     sequence $S_0$ and $N-n+k+(l-m)$ viruses in other sequences.
     Thus there are $N+k+l$ sequences in $S_i$ for $1 \le i$.  In the next step, we will uniformly remove $l+k$ sequences
     so that the total population is still $N$.

 \vskip 2 mm

\noindent C. We randomly choose $h$ elements from a pool of $l+k$
elements in $S_0$ and each element is chosen
 independently with the probability $x$. Thus the probability
 to remove $h$ elements from $S_0$ is:
\begin{eqnarray}
\label{e14}\rho_4(h)=\frac{x^h(1-x)^{(l+k-h)}}{h!(l+k-h)!}.
\end{eqnarray}
Besides, we remove $l+k-h$ elements from $S_i$ for $i
>0$. We have that during one iteration step $n \to n+m-h$,
therefore we need to find the average $<e^{-\phi'(m-h)}>$ via the
distributions $\rho_1(l)\rho_2(k)\rho_3(m)\rho_4(h)$. We consider:
\begin{eqnarray}
\label{e15}<e^{-\phi'(m-h)}>= \nonumber\\
\sum _{l,k,m,h} \frac{n!(U A)^l(1-U A)^{n-l}}{l!(n-l)!}
\frac{l!Q^m(1-Q)^{l-m}e^{-\phi'm}}{m!(l-m)!}\times\nonumber\\
\frac{(N-n)!U^k(1-U)^{N-n-k}}{k!(N-n-k)!}
\frac{(l+k)!x^h(1-x)^{(l+k-h)}}{h!(l+k-h)!}e^{\phi'h}.
\end{eqnarray}
First we transform
\begin{eqnarray}
\label{e16} \sum _m
\frac{l!Q^m(1-Q)^{l-m}e^{-\phi'm}}{m!(l-m)!}=(Qe^{-\phi'}+1-Q)^l.
\end{eqnarray}

Using the transformation
\begin{eqnarray}
\label{e17} \sum _{h}
\frac{(l+k)!x^h(1-x)^{(l+k-h)}}{h!(l+k-h)!}e^{\phi'h}=(1-x+xe^{\phi'})^{l+k},
\end{eqnarray}
we obtain
\begin{eqnarray}
\label{e18} \sum _{l,k} \frac{n!(U A)^l(1-U A)^{n-l}}{l!(n-l)!}
\frac{(N-n)!U^k(1-U)^{N-n-k}}{k!(N-n-k)!}\times\nonumber\\
(Qe^{-\phi'}+1-x)^l(1-x+xe^{\phi'})^{l+k}.
\end{eqnarray}
The sum over $l,k$ gives  an equation
\begin{eqnarray}
\label{ce19} \frac{d\phi}{dt}=F(\phi'),
\end{eqnarray}
where
\begin{eqnarray}
\label{e20}
F(\phi')&=&xA[(Qe^{-\phi'}+1-Q)(xe^{\phi'}+1-x)-1]\nonumber\\
&+&x(1-x)(e^{\phi'}-1).
\end{eqnarray}
We need to consider the first two terms in the $\phi'$ expansion
\begin{eqnarray}
\label{e21}
F(\phi')\approx -x[(QA-1)-(A-1)x]\phi'\nonumber\\
+x[QA(1-2x)+(A-1)x+1]\frac{\phi'^2}{2}.
\end{eqnarray}
In the steady state we consider $F(\phi')=0$. We expand Eq.(21) in
powers of $y\equiv x-\frac{(QA-1)}{(A-1)}$ and find the following
steady state solution:
\begin{eqnarray}
\label{e22}
\phi'=-2\frac{(A-1)y}{Q(1-Q)\frac{2A^2}{(A-1)}-(2QA+1-A)y}.
\end{eqnarray}
Therefore,
\begin{eqnarray}
\label{e23} \phi''(0)=-\frac{(A-1)^2}{Q(1-Q)A^2},
\end{eqnarray}
and eventually we obtain for the variance $V$ of distribution
\begin{eqnarray}
\label{e24} V=N <y^2>\equiv
N(<p_0^2>-<p_0>^2)=\frac{Q(1-Q)A^2}{(A-1)^2}.
\end{eqnarray}
In Appendix C, we derive the steady state distribution and the
variance for the  Eigen model with degradation Eq. (C6).

 \section{Discussion}

 The investigation of finite population problem is the hardest
mathematical problem in evolution theory.  In this article we solved
exactly some aspects of finite population version of Crow-Kimura and
Eigen model with degradation. We calculated the variance of the
distribution for the mean fitness in the equlibrium. Our equation
could be applied to calculate the dynamics of the distribution as
well.

The quasispecies model, especially the one with single peak fitness
and its simple generalizations, has a lot of applications in the
virus evolution \cite{ei02}, cancer modeling \cite{so04} and
molecular evolution \cite{wo11}. Therefore any rigorous results here
should be welcomed.

In this article we considered just one aspect of convergence of
finite population result to the results in infinite population
considering the variance of the mean fitness. According this
criteria, $N\sim L^2$ is large enough to have the same mean fitness
as the infinite population with the accuracy $1/L$. Actually an
important open problem is to investigate the equilibrium here
(mutation-selection), like the equilibrium in the thermodynamics,
and how the equilibrium is affected by finite size of population.

\begin{figure}
\centerline{\includegraphics[width=0.6\columnwidth]{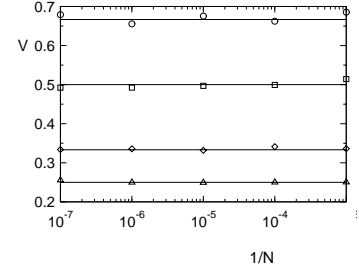}}
\caption{Verification of the Equation (\ref{e10}) for variance:
$V=1/J$. The horizontal lines from top from bottom are analytic
results for $J=1.5,~2,~3$ and $4$, respectively. Circles, squares,
diamonds, and triangles represent numerical data for $J=1.5,~2,~3$
and $4$, respectively. Analytic and numerical results are quite
consistent with each other.} \label{fig2}
\end{figure}

 \acknowledgments
DBS thanks U.S. Defense Advanced Research Projects Agency DARPA
grant No. HR00110510057, DARPA Prophecy Program and Academia Sinica
for financial support. This work was supported in part by NSC
100-2112-M-001-003-MY2 and NCTS in Taiwan.

\section{Appendix A. Crow-Kimura model}

 \setcounter{equation}{0}
\renewcommand{\theequation}{A.\arabic{equation}}

We here consider the infinite population model.
 The $M \equiv 2^L$ genome configurations (sequences) are defined
as chains of $L$ spins $s_k,~1\le k \le L$ having values $\pm 1$.
There is a reference sequence $S_0$ with all spins $+1$. We define
the Hamming distance between the given sequence and the reference
sequence by
 $\sum_k(1-s_k)/2 \equiv N(1-m)/2$, where $m$ is an overlap.

  The state
of system is specified by the $M$ relative frequencies $p_i,0 \le i
\le M-1$:
\begin{eqnarray}
\label{a1}
\frac{{dp}_i}{dt}=\sum_jA_{ij}p_j-p_i\sum_jp_j r_j\nonumber\\
A_{ij}=\delta_{ij} r_j+m_{i j}.
\end{eqnarray}
Here $m_{ij}$ is the rate of mutation from the sequence $j$ to the
sequence $i$,and $ r_{i}$ is the fitness. Two sequences  have a
Hamming distance $d_{ij}=(L-\sum_k s_i^k s_j^k)/2$.
 Here $m_{ii}=-\gamma_0 $. When $d_{ij}=1$, $m_{ij}=\gamma_0/L$,
$m_{ij}=0$ for $d_{ij}>1$ \cite{ba97}. We are interested in the
symmetric fitness landscape with
\begin{eqnarray}
\label{a2}  r_i=f(1-2d_{i0}/L).
\end{eqnarray}
We choose the first $L+1$ sequences such that the $l$-th sequence
has Hamming distance $l$ from the reference sequence, $0\le l\le L$.
Then our $ r_l$ are connected with the $\hat r_l$ in the main text
as
\begin{eqnarray}
\label{a3} \hat r_l=\exp[ r_l U].
\end{eqnarray}
where
\begin{eqnarray}
\label{a4}  U=\tau \gamma_0,
\end{eqnarray}
where $\tau$ is the iteration duration  and $l$ in Eq.(A3) is the
Hamming class of the $i$-th sequence.
 In the main text  we consider discrete time evolution with
minimal time interval $\tau=U$, choosing $\gamma_0=1$.

\renewcommand{\theequation}{B.\arabic{equation}}
\setcounter{equation}{0}
\section{Appendix B. Eigen model}

In Eigen's quasispecies theory \cite{ei71,ei89}, the $i$-th sequence
produces offspring of the type $j$ with the probability
$Q_{ij}=q^{L-d_{ij}}(1-q)^{d_{ij}}$, where $1- q$ is the average
number of errors per site and $d_{ij}$ is the Hamming distance .

Eigen proposed that $p_i$ satisfy \cite{ei71,ei89}

\begin{eqnarray}
\label{c1} \frac{dp_i}{dt} &=& \{Q_{ii} r_i -D_i \sum_{k}\hat r_k
p_k(t)\}p_i(t) \nonumber \\
&+& \sum_{k\ne i}Q_{ik} r_kp_k(t).
\end{eqnarray}
Here $D_i$ describes degradation. It is convenient to work with the
error rate $\gamma\equiv L(1- q)$, leading to $Q=e^{-\gamma}$.

\renewcommand{\theequation}{C.\arabic{equation}}
\setcounter{equation}{0}
\section{Appendix C. Eigen model with degradation}

We now consider the Eigen model when there is degradation $D$ in the
peak sequence $S_0$, and zero degradation for other sequences $S_i$
for $i>0$. In this case we should add the random sampling for the
degradation case. The calculation procedure is similar to the case
in the section of Eigen model. We should just modify the iteration
sub-steps from that section, after the point B.

C. There is a dilution of the population with the peak sequence. We
randomly choose $t$ elements from a pool of $n$ elements and each
element is chosen independently with the probability $U D$.

D. There is a dilution of the whole population, keeping strictly the
total population size as $N$.

We randomly choose $h$ elements from a pool of $l+k-t$ elements and
each element is chosen independently with the probability $U(1-x)$.

Now after one iteration step $n\to n+m-h-t$. Thus we should
calculate $<e^{-\phi'(m-t-h)}>$. We get the following expression:
\begin{eqnarray}
\label{d1}
<e^{-(m-t-h)\phi'}>=\nonumber\\
\sum_{l,k,m,t,h}\frac{n!(U A)^l(1-U A)^{n-l}}{l!(n-l)!}\frac{l!Q^m(1-Q)^{l-m}e^{-\phi'm}}{m!(l-m)!}\times\nonumber\\
\frac{n!e^{\phi't}e^{\phi'h)}(U D)^t(1-U D)^{n-t}}{t!(n-t)!}\times\nonumber\\
\frac{(N-n)!U^k(1-U)^{N-n-k}}{k!(N-n-k)!}\frac{(l+k-t)!x^k(1-x)^{(l+k-t-h)}}{h!(l+k-t-h)!}
\end{eqnarray}

We first perform the sum over $h$:
\begin{eqnarray}
\label{d2}
 \sum_h\frac{(l+k-t)!x^h(1-x)^{(l+k-t-h)}}{h!(l+k-t-h)!}e^{\phi'h}\nonumber\\
=(1+x(e^{\phi'}-1))^{l+k-t},
\end{eqnarray}
then perform the sum over $t$:
\begin{eqnarray}
\sum_t\label{d3} \frac{n!e^{\phi't}(U D)^t(1-U
D)^{n-t}}{t!(n-t)!}(1+x(e^{\phi'}-1))^{-t}=\nonumber\\
 (U\frac{de^{\phi'}}{(1+x(e^{\phi'}-1))}+1-U)^{Nx} \nonumber\\=
\exp[U x(\frac{e^{\phi'}}{1+x(e^{\phi'}-1)}-1)dN].
\end{eqnarray}
Comparing our formulas with the expression of $F(\phi')$ from the
section of Finite population Eigen model, we find just new
additional term to those of Eq.(20). Eventually we have:
\begin{eqnarray}
\label{D4} \frac{d\phi'}{dt}
= xA [(Qe^{-\phi'}+1-Q)(xe^{\phi'}+1-x)-1]\nonumber\\
+ x(\frac{e^{\phi'}}{1+x(e^{\phi'}-1)}-1)D+(1-x)x(e^{\phi'}-1).
\end{eqnarray}
We expand in powers of $\phi'$:
\begin{eqnarray}
\label{D5} F(\phi')
\approx -[(QA-1-D)-(A-1-D)x]\phi'+\nonumber\\
\frac{\phi'^2}{2}[QA(1-2x)+(A-1)x+1+D(1-x)(1-2x)].
\end{eqnarray}
Putting the value of $x=\frac{AQ-D-1}{A-D-1}$, we derive
$$F(\phi')\approx
 -[(QA-1-D)-(A-1-d)x]\phi'$$
$$+\frac{\phi'^2}{2}\frac{(2 a (-1 + Q)((D + D^2 + (-1 +a) a
Q- 2 a D Q)))}{(1+D-a)^2}.$$
and obtain for the variance $V$
\begin{eqnarray}
\label{D7} V&=&\frac{A(1-Q)((A-1)AQ+2AQd-d-d^2)}{(A-1-d)^3}.
\end{eqnarray}
For $D=0$, Equation (\ref{D7}) reduces to Equation (\ref{e20}) for
the Eigen model without degradation.

\end{document}